\documentclass[conference]{IEEEtran}
\IEEEoverridecommandlockouts
\usepackage{todonotes}
\usepackage{cite}
\usepackage{hyperref}
\usepackage{amsmath,amssymb,amsfonts}
\usepackage{algorithmic}
\usepackage{graphicx}
\usepackage{textcomp}
\usepackage{enumerate}
\usepackage{xcolor}
\usepackage{subcaption}
\usepackage{booktabs}
\def\BibTeX{{\rm B\kern-.05em{\sc i\kern-.025em b}\kern-.08em
    T\kern-.1667em\lower.7ex\hbox{E}\kern-.125emX}}
\begin{document}

\title{Towards Comparing Performance of Algorithms in Hardware and Software 
\thanks{This work is supported by the Innovation Fund Denmark for the project DIREC (9142-00001B) and by CERCIRAS Cost Action CA19135.}
}

\author{\IEEEauthorblockN {Maja H.\ Kirkeby}
\IEEEauthorblockA{\textit{Dept. of People and Technology} \\
\textit{Roskilde University,}\\
Roskilde, Denmark \\
kirkebym@acm.org}
\and
\IEEEauthorblockN{Martin Schoeberl}
\IEEEauthorblockA{\textit{DTU Compute} \\
\textit{Technical University of Denmark}\\
Lyngby, Denmark \\
masca@dtu.dk}
}

\maketitle

\begin{abstract}
In this paper, we report on a preliminary investigation of the potential performance gain of programs implemented in field-programmable gate arrays (FPGAs) using a high-level language Chisel compared to ordinary high-level software implementations executed on general-purpose computers and small and cheap computers. 
FPGAs inherently support parallel evaluations, while sequential computers do not. For this preliminary investigation, we have chosen a highly parallelizable program as a case study to show an upper bound of performance gain. The purpose is to demonstrate whether or not programming FPGAs has the potential for performance optimizations of ordinary programs.

We have developed and evaluated Conway's Game of Life for an  FPGA, a small and cheap computer Raspberry Pi 4, and a MacBook Pro Laptop.
We have compared the performance of programs over different input sizes to decide the relative increase in runtime.

\end{abstract}

\begin{IEEEkeywords}
Time consumption, FPGA, Raspberry Pi, General-purpose computers
\end{IEEEkeywords}

\section{Introduction}
Using FPGAs is known to improve the performance when programmed in the hardware specifying languages such as Verilog or VHDL~\cite{Zhou2020}. However, these languages are oriented toward engineers and require highly specialised knowledge. Therefore, alternative approaches are interesting, such as using FPGAs as accelerators for C/C++ programs~\cite{Jmaa2019} or, as in this study, where we use the programming language Chisel, which is oriented toward software developers. 

This paper intends to indicate an upper bound for how much Chisel implementations could improve typical implementations of software programs. 
Investigating how accessible these approaches are for software developers is beyond the scope of this paper. Instead, to emulate the level of language familiarity will use a naive solution written in Chisel and compare it to a naive solution written in Java; according to Stack overflows yearly reports, it has been in the top 5 most popular programming languages at least since 2017~\footnote{
\url{https://insights.stackoverflow.com/survey/2017}
\url{https://insights.stackoverflow.com/survey/2018}, \url{https://insights.stackoverflow.com/survey/2019},
\url{https://insights.stackoverflow.com/survey/2020},
\url{https://insights.stackoverflow.com/survey/2021}}. 

FPGA units can support or replace different devices depending on the algorithm. Therefore we have chosen to compare the performance of the same algorithm running on an FPGA with different kinds of computers: a typical general-purpose computer in the form of a MacBook Pro and a small and cheap computer, exemplified by a Raspberry Pi.

Because we look for upper bounds, we choose a highly parallelizable software algorithm, and we limit the size of the problems it solves to fit into the FPGA without having to read and write to memory. An example of such an algorithm program is Conway's Game of Life~\cite{berlekamp1982life}. The Game of Life is an algorithm which exemplifies how complex behaviour can arise from simple rules, e.g., it has been used for constructing a Turing machine~\cite{Rendell2016}.

In the following, we introduce the specifications of Conway's Game of Life and the different implementations that we use for comparison. Section~III describes our experimental setup. Section~IV evaluates the performance, and Section~V discusses the setup, the resource utilization, and a conservative estimation of the energy consumption of our experiment. Section~VI concludes the paper and provides an outlook for future work.

\section{Conway's Game of Life} \label{sec:Game of Life}
There are many variations\footnote{See \url{http://conwaylife.com/}.} of Conway's Game of Life~\cite{berlekamp1982life}, which is a zero-player game defined on cellular automata. In this report, we define Game of Life as follows. Cellular automata are 2D-grids, called worlds, where each grid-cell has eight neighbours, and each cell can have one of two states: dead or alive. For each time step, the cell states are updated according to their own and their neighbouring cells' states from the previous time step.
\begin{enumerate}[Rule 1:]
\item Any live cell with two or three live neighbours survives.
\item Any dead cell with three live neighbours becomes a live cell.
\item All other live cells die in the next generation. Similarly, all other dead cells stay dead.
\end{enumerate}
The initial state is given as input to the program. Because each cell depends only on nearby cells, Game of Life is highly parallelizable.
An example with two time steps in Game of Life can be seen in Figure~\ref{fig:Beacon}; the figure displays three 6x6 worlds where "O" are live cells and "." are dead cells. The left-most grid~(\ref{fig:beacon-0}) is the initial world at time step $0$, the middle grid~(\ref{fig:beacon-1}) is the updated world at time step 1 and the right-most~(\ref{fig:beacon-2}) is the world at time step 2. The world at timestep 0 and 2 are equal, and patterns with this particular type of behaviour are oscillators. 

\begin{figure}[t!]
\centering
\begin{subfigure}[t]{0.1\textwidth}
\begin{verbatim}
	......
	.OO...
	.O....
	....O.
	...OO.
	......
\end{verbatim}
\caption{}\label{fig:beacon-0}
\end{subfigure}%
~\begin{subfigure}[t]{0.1\textwidth}
 \begin{verbatim}
	......
	.OO...
	.OO...
	...OO.
	...OO.
	......
\end{verbatim}
\caption{}\label{fig:beacon-1}
\end{subfigure}%
~\begin{subfigure}[t]{0.1\textwidth}
 \begin{verbatim}
	......
	.OO...
	.O....
	....O.
	...OO.
	......
\end{verbatim}
\caption{}\label{fig:beacon-2}
\end{subfigure} 

    \caption{Beacon: Oscilator}
    \label{fig:Beacon}
\end{figure}

\subsection{Software Implementation in Java}
We have implemented the Game of Life in Java using 2D arrays to represent a world. As on optimization, we copy the initial array that contains the starting world into an array that is two rows and two columns larger, where the border cells are dead; this avoids testing for boundary cases when counting live neighbours. 
For comparison of the implementations, we have included the code for computing the next state for a single cell at position (i,j); the array "curr" contains the cell's value at the current time step and the array "next" contains the cell's value at the next time step. 
\begin{verbatim}
L01  int cnt = 0;
L02  for (int k = -1; k <= 1; ++ k) {
L03      cnt += curr[i-1][j+k];
L04      cnt += curr[i+1][j+k];
L05  }
L06  cnt += curr[i][j-1];
L07  cnt += curr[i][j+1];
L08  if ((curr[i][j] == 1 && cnt == 2) || 
L09      (cnt == 3)) {
L10      next[i][j] = 1;
L11  } else {
L12      next[i][j] = 0;
L13  }
\end{verbatim}
First, it counts the number of neighbouring cells alive in the previous step and saves it into the \verb+cnt+ (line \verb+L01+-\verb+L07+). Afterwards, the rules of Game of Life are applied using a single if-expression (line \verb+L08+-\verb+L13+).

\subsection{Hardware Implementation in Chisel}

We have implemented the Game of Life in hardware, using the hardware construction language Chisel~\cite{chisel:dac2012, chisel:book}. Each cell consists of a single D flip-flop, a population counter of the neighbours, and the conditions for staying alive, dying, or generating a live cell. This simple cell structure is replicated for the given world size. 
The registers are reset according to the starting pattern of the world.

The inputs of the boundary cells are connected to 0 at hardware construction time. Therefore, those conditions do not need to be evaluated at run time. This is a similar optimization to increasing the array size for the software implementation.

The Game of Life hardware is embarrassingly parallel, as all cells \emph{execute} in parallel. Each \emph{execution} is a very simple function. Therefore, we expect that this circuit can operate at a high frequency. The whole world is updated with each clock cycle.

To demonstrate the conciseness of hardware described in Chisel, we show the full code for a single cell:
\begin{verbatim}
L01  val regCell = RegInit(init.B)
L02  val cnt = PopCount(io.in)
L03
L04  when((regCell && cnt === 2.U) ||
L05    cnt === 3.U) {
L06    regCell := true.B
L07  } .otherwise {
L08    regCell := false.B
L09  }
\end{verbatim}
\verb'RegInit(init.B)' defines a hardware register for saving the cell's value between time steps, i.e., it has the cell's value from the last time step (line \verb+L01+). \verb'PopCount(io.in)' which is given the neighbouring cells' bit values from the previous time step as input and returns the number of bit values that are 1 (line \verb+L02+). Afterwards, it implements the rules of Game of Life similarly to the software implementation ((line \verb+L04+-\verb+L09+). Chisel is embedded in Scala, and the `\verb+.B+' and `\verb+.U+' are constructors that cast Scala values into Chisel types Bool and Unsigned integers deciding a suitable number of bits.

\section{Experimental Setup}
In the previous section, we described the algorithm and the implementations, and in this section, we describe the setup for measuring their performance on different devices. FPGA units can support or replace various types of devices depending on the algorithm, and the systems may have different computation power. Thus, for comparison, we have chosen to evaluate the performance of both a typical general-purpose computer and a small and cheap computer.
\begin{figure*}[!t]
\centering

\includegraphics[width=0.3\textwidth]{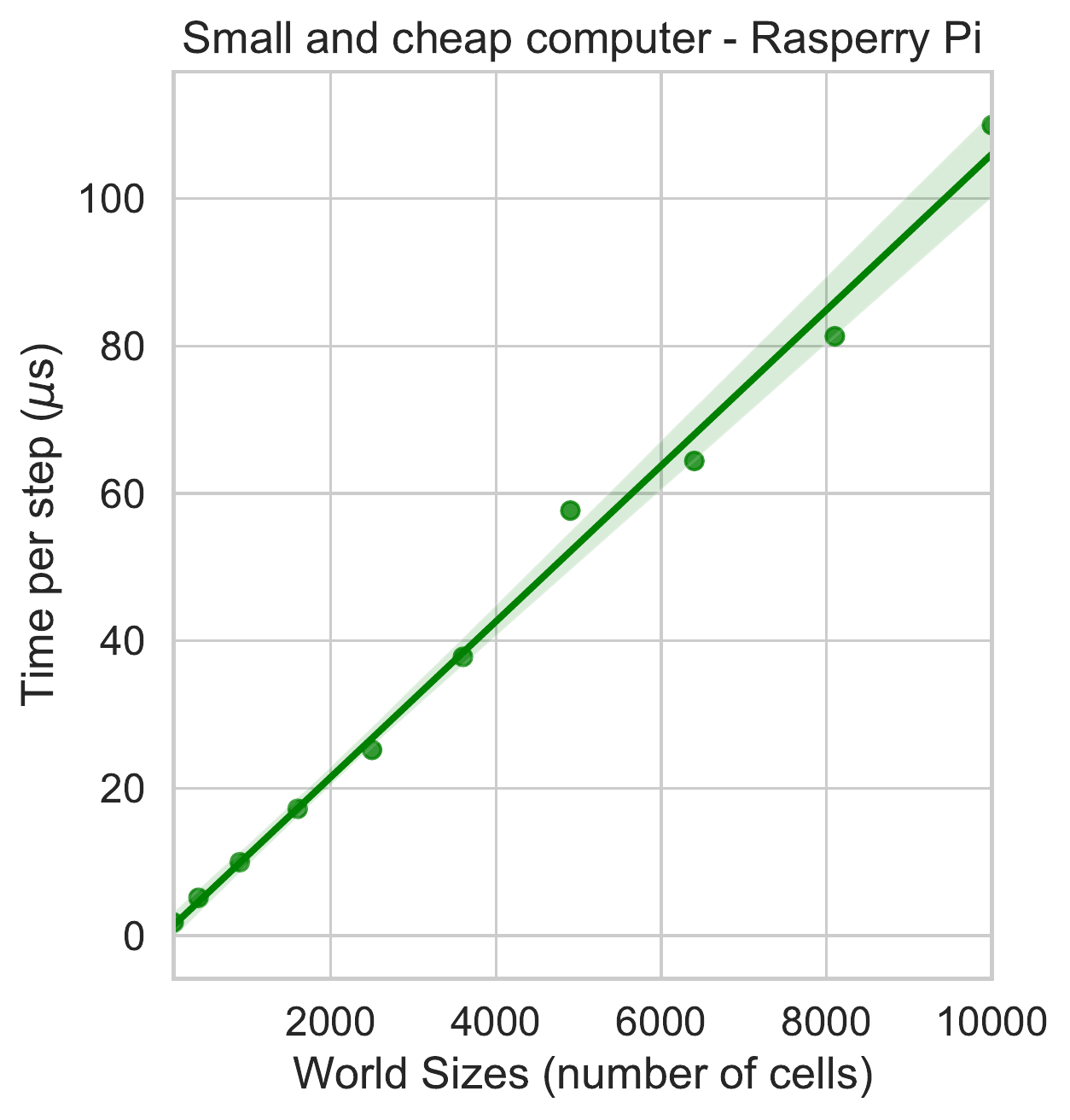}~%
\includegraphics[width=0.3\textwidth]{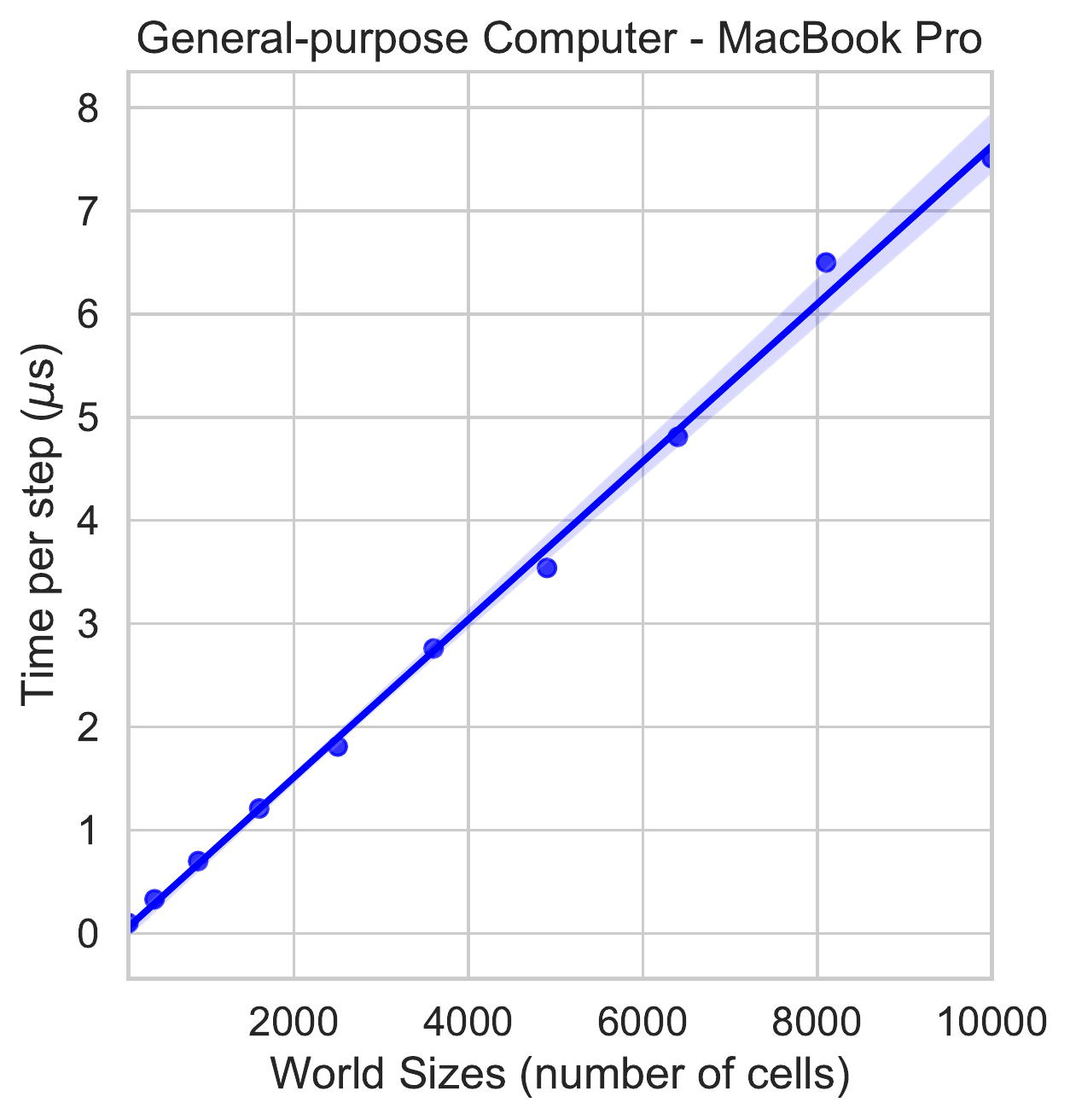}~%
\includegraphics[width=0.3\textwidth]{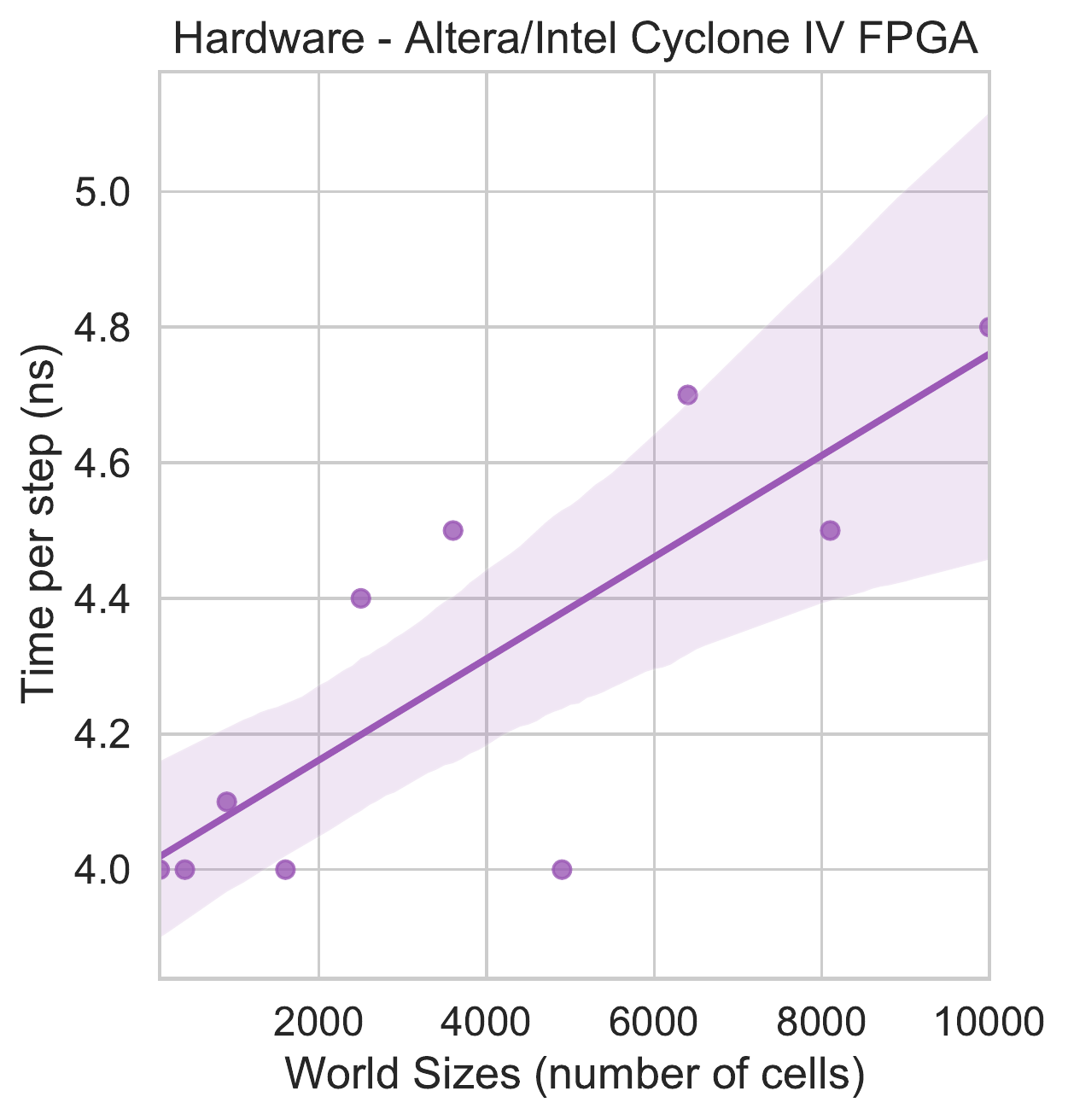}
\caption{The trend lines (linear regression) of time pr step (in us and ns) over increasing world sizes: 10x10, 20x20, $\ldots$, 100x100.}
    \label{fig:results}
\end{figure*}
\subsection{Hardware}
In the following experiments, we have compared the execution times for the implementations described in the previous section when executed on three devices. The Java implementation is executed on a general-purpose computer and a small and cheap computer, and the hardware implementation is executed on an FPGA board with the following specifications:
\begin{itemize}
    \item Typical general-purpose computer: MacBook Pro with a 2,6 GHz 6-Core Intel Core i7 processor and 16 GB memory.
    \item Small and cheap computer: Raspberry Pi 4 Model B 4GB RAM with 1.5GHz with a Broadcom BCM2711, Quad core Cortex-A72 (ARM v8) 64-bit SoC.
    \item Hardware: a Altera/Intel Cyclone IV FPGA found on the DE2-115 evaluation board.
\end{itemize}

\subsection{Inputs}
To answer how the data input size influences execution time, we evaluate the execution time for random worlds with different world sizes: 10x10, 20x20, $\ldots$, 100x100. For easy comparison, we consider the average execution time per time step, and we execute at least 1000000 time steps to obtain robust results.

\section{Performance}
\begin{figure}[b]
    \centering
    \includegraphics[width=0.5\textwidth]{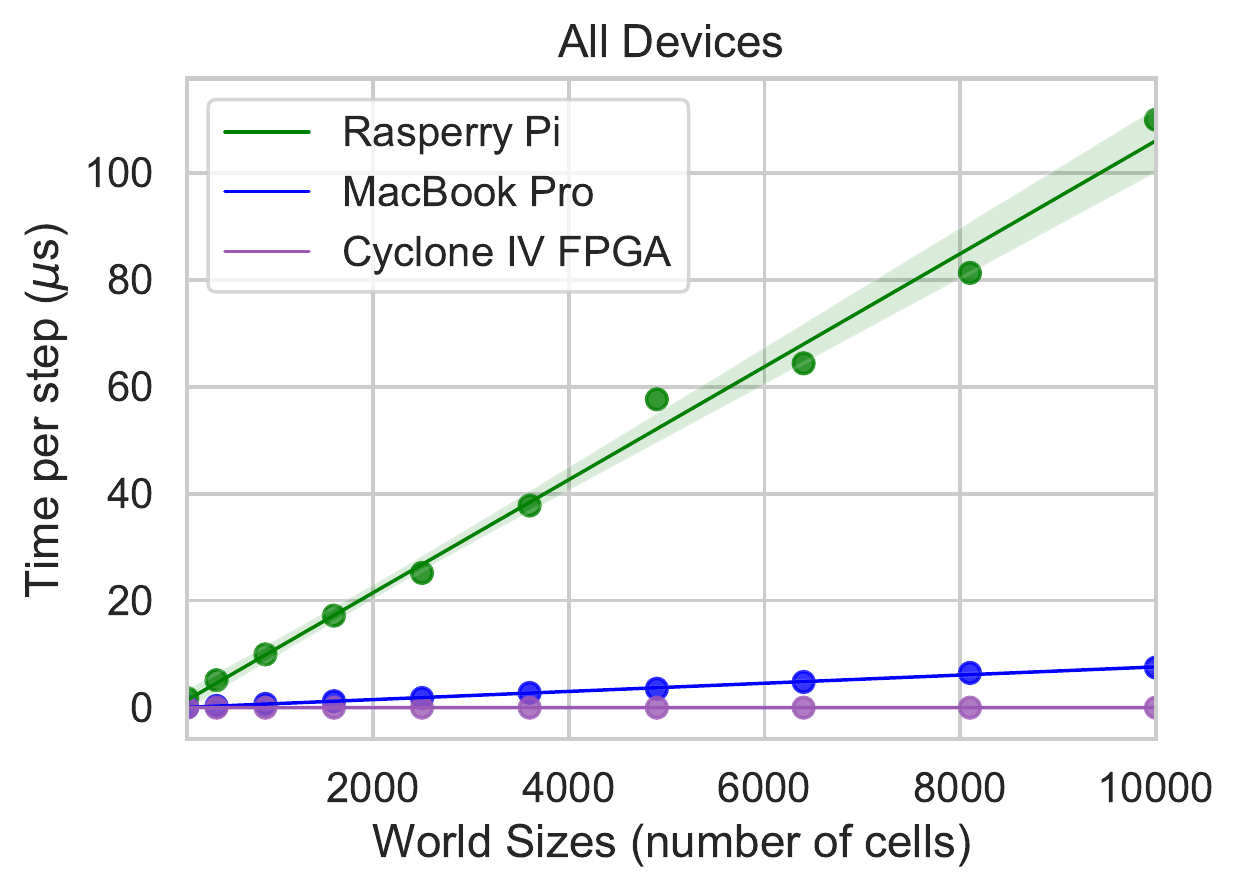}~%
    \caption{Comparison of time pr step (us) over increasing world sizes: 10x10, 20x20, $\ldots$, 100x100 for the small computer, typical general-purpose computer and hardware implementation.}
    \label{fig:results-overview}
\end{figure}

We report the average execution times for a single Game of Life iteration for increasing world sizes, i.e., the average execution time for computing the next world in the Game of Life. The Figure~\ref{fig:results} depicts -for each of the three devices- the execution time and their trend lines (linear regression). For all types of devices, the trend lines show an increase over the world sizes; however, their growth rates are very different, as shown in Figure~\ref{fig:results-overview}.

\begin{table}[b]
  \centering
  \caption{The execution time in us and speed-up of FPGA over software executions.}
\begin{tabular}{crrrrrr}
    \toprule
    & &\multicolumn{3}{c}{Execution time pr step (us)} & \multicolumn{2}{c}{FPGA Speedup}\\
World & Cells&Mac&Raspberry&FPGA& Mac& Raspberry\\
\midrule
10x10 & 100&0.10&1.783&0.0040&25&445\\
20x20 & 400&0.33&5.137&0.0040&82&1284\\
30x30 & 900&0.70&9.965&0.0041&170&2430\\
40x40 & 1600&1.21&17.212&0.0040&302&4302\\
50x50 & 2500&1.81&25.204&0.0044&411&5728\\
60x60 & 3600&2.76&37.822&0.0045&613&8404\\
70x70 & 4900&3.54&57.665&0.0040&884&14416\\
80x80 & 6400&4.81&64.396&0.0047&1023&13701\\
90x90 & 8100&6.50&81.309&0.0045&1444&18068\\
100x100 & 10000&7.51&109.964&0.0048&1564&22909\\
\bottomrule
\end{tabular}
    \label{tab:speedup}
\end{table}

Table~\ref{tab:speedup} shows the average execution time for a single time step and the speedup provided by the FPGA implementation compared to the Java software implementation executed on a MacBook Pro and the Raspberry Pi. From these results, we see that the gain in performance increase with the measured world sizes; in Section~\ref{sec:Discussion} we comment on the limits of these results.

\section{Discussion}\label{sec:Discussion}
The presented results show that the FPGA implementation of a 40x40 Game of Life world is around 630 times faster than the software implementation in Java when running on a ten times higher clocked general purpose processor. Note that the Altera/Intel Cyclone IV FPGA we are using in this experiment is a relatively old FPGA.
With a newer FPGA, we expect to be able to implement larger worlds and execute them at a higher clock frequency.

When increasing the problem size to 100 x 100, the performance difference between the hardware solution and the software implementation increases linearly. However, this will not continue when the hardware implementation cannot be expressed within the number of logic elements available on the FPGA board. If we were to handle larger world sizes, the hardware implementation must be redesigned, whereas the software implementations do not encounter this issue.

While the Game of Life is an artificial workload, its parallelizable nature made it an ideal candidate for indicating an upper bound for speedups when moving algorithms from software into an FPGA.

\subsection{Resource Utilization}
In the following, we consider the hardware implementation's resource utilization and estimate the impact on energy consumption for running Game of Life on the small computer and the hardware.

We have implemented the Game of Life of different sizes in an Altera/Intel Cyclone IV FPGA found on the DE2-115 evaluation board.
We report the size of the design in logic elements (LEs) and registers.
An LE represents one 4-bit lookup table.
For synthesize, we used the Quartus 19.1.0 Lite Edition.

Table~\ref{tab:fpga} shows the FPGA implementation's resource consumption for different world sizes.
We can see that the size grows linear. The maximum frequency of the circuit is reported between 209~MHz and 250~MHz. Therefore, when we assume running it at 200~MHz, we can compute one iteration in 5~ns.

\begin{table}
  \centering
  \caption{The resource utilization and minimum iteration time of different sized Game of Life worlds in an FPGA.}
  \begin{tabular}{rrrc}
    \toprule
    Size & LEs & Registers & min. Clock Period \\
    \midrule
10 x 10 & 804 & 104 & 4.0 ns\\
20 x 20 & 3539 & 404 & 4.0 ns \\
30 x 30 & 7995 & 904 & 4.1 ns \\
40 x 40 & 14463 & 1604 & 4.0 ns \\
50 x 50 & 23439 & 2504 & 4.4 ns \\
60 x 60 & 34414 & 3604 & 4.5 ns \\
70 x 70 & 45119 & 4904 & 4.0 ns \\
80 x 80 & 59136 & 6404 & 4.7 ns \\
90 x 90 & 75102 & 8104 & 4.5 ns \\
100 x 100 & 97871 & 10004 & 4.8 ns \\
    \bottomrule 
  \end{tabular} 
 \label{tab:fpga}
\end{table}

As expected, we use one register per cell. However, the number of LEs per cell is surprisingly high, an average of around 9 LEs per cell. We assume that the Chisel PopCount method has some room for improvement. However, as we aim for a method that enables software developers to describe their algorithms in hardware, we are avoiding optimization tricks.

\subsection{Estimated Energy Consumption}
In this experiment, we did not measure the power or energy consumption of the FPGA implementation. However, the DE2-115 FPGA board comes with a power supply of 24~W. Therefore, this is the upper bound of power consumption of the whole FPGA board, including peripheral devices and external memories.

For a conservative estimate, we assume those 24~W as an upper bound on the power consumption and an operating frequency of 200~MHz; then, one iteration of a 100 x 100 Game of Life world consumes 96~nJ.

In comparison, the Raspberry Pi has been reported to consume average 6.4~W when all four cores are busy\footnote{\url{https://www.pidramble.com/wiki/benchmarks/power-consumption}} and one iteration of a 100x100 world takes 0.109964 ms. Thus, a conservative estimate of the energy consumption for one iteration of a 100x100 world is 0.703769~mJ. 

From this conservative comparison, it seems that the hardware implementation can considerably improve energy consumption compared to the Raspberry Pi 4.  

\section{Conclusion and Future work}
In this paper, we have demonstrated that a hardware implementation using Chisel compared to software implementations using Java can provide severe performance optimizations for highly parallelizable problems and where the hardware implementation can avoid reading and writing from memory.
Therefore, we have established an upper bound of performance gain and energy reduction when using an FPGA compared to a general-purpose computer or a small and cheap computer.

In the future, we will extend this research to:  (i) Comparing the parallel hardware implementations' energy consumption and performance with parallel programs running on GPUs, (ii) Studying also often used algorithms, such as path-finding algorithms or sorting algorithms, and (iii) Measuring the energy consumption of the different implementations of the algorithms in order to study the changes in energy consumption and execution time.

\bibliography{bibliography.bib}
\bibliographystyle{plain}

\end{document}